\documentclass[12pt]{article}
\usepackage[nosort]{cite}
\usepackage{graphicx}
\usepackage[font={scriptsize,singlespacing}]{caption}
\usepackage{epstopdf}
\usepackage{amsmath}
\usepackage{amssymb}
\usepackage{subfigure}
\usepackage{hyperref}
\usepackage{url}
\usepackage{xcolor}
\usepackage{color}

\def\sideremark#1{\ifvmode\leavevmode\fi\vadjust{\vbox to0pt{\vss
 \hbox to 0pt{\hskip\hsize\hskip1em
 \vbox{\hsize2cm\tiny\raggedright\pretolerance10000
 \noindent #1\hfill}\hss}\vbox to8pt{\vfil}\vss}}}%
\definecolor{amaranth}{rgb}{0.9, 0.17, 0.31}
\definecolor{purple(munsell)}{rgb}{0.62, 0.0, 0.77}
\definecolor{americanrose}{rgb}{1.0, 0.01, 0.24}
\definecolor{palatinateblue}{rgb}{0.15, 0.23, 0.89}
\definecolor{royalblue(web)}{rgb}{0.25, 0.41, 0.88}
\definecolor{hanpurple}{rgb}{0.32, 0.09, 0.98}
\definecolor{beaublue}{rgb}{0.74, 0.83, 0.9}
\definecolor{carminered}{rgb}{1.0, 0.0, 0.22}
\definecolor{brightpink}{rgb}{1.0, 0.0, 0.5}
\hypersetup{ linktoc=all,
    colorlinks, linkcolor={palatinateblue},
    citecolor={brightpink}, urlcolor={purple(munsell)}
}

\begin{document}
\thispagestyle{empty}
\begin{center}

\null \vskip-1truecm \vskip2truecm

{\Large{\bf \textsf{Generalized Uncertainty Principle:}}}
\vskip0.1truecm
{\Large {\bf\textsf{
Implications for Black Hole Complementarity}}}

\vskip1truecm
\textbf{\textsf{Pisin Chen}}
{\footnotesize\textsf{\\(1) Leung Center for Cosmology and Particle Astrophysics \& \\ Graduate Institute of Astrophysics \& Department of Physics,\\  National Taiwan University,
Taipei 10617, Taiwan\\ (2) Kavli Institute for Particle Astrophysics and Cosmology, \\SLAC National Accelerator Laboratory, Stanford University, CA 94305, U.S.A}\\
{\tt Email: chen@slac.stanford.edu}}\\

\vskip0.4truecm
\textbf{\textsf{Yen Chin Ong}}\\
{\footnotesize \textsf{(1) Leung Center for Cosmology and Particle Astrophysics,  \\National Taiwan University,
Taipei 10617, Taiwan\\
(2) Nordita, KTH Royal Institute of Technology and Stockholm University, \\ Roslagstullsbacken 23,
SE-106 91 Stockholm, Sweden}\\
{\tt Email: ongyenchin@member.ams.org}}\\

\vskip0.4truecm
\textbf{\textsf{Dong-han Yeom}}\\
{\footnotesize \textsf{(1) Yukawa Institute for Theoretical Physics,\\ Kyoto University, Kyoto 606-8502, Japan\\
(2) Center for Quantum Spacetime, Sogang University,\\ Seoul 121-741, Republic of Korea}\\
{\tt Email: innocent.yeom@gmail.com}}\\

\end{center}
\vskip1truecm \centerline{\textsf{ABSTRACT}} \baselineskip=15pt

\medskip

At the heart of the black hole information loss paradox and the firewall controversy lies the conflict between quantum mechanics and general relativity. Much has been said about quantum corrections to general relativity, but much less in the opposite direction. It is therefore crucial to examine possible corrections to quantum mechanics due to gravity. Indeed, the Heisenberg Uncertainty Principle is one profound feature of quantum mechanics, which nevertheless may receive correction when gravitational effects become important. Such generalized uncertainty principle [GUP] has been motivated from not only quite general considerations of quantum mechanics and gravity, but also string theoretic arguments. We examine the role of GUP in the context of black hole complementarity. We find that while complementarity can be violated by large $N$ rescaling if one assumes only the Heisenberg's Uncertainty Principle, the application of GUP may save complementarity, but only if certain $N$-dependence is also assumed.  This raises two important questions beyond the scope of this work, i.e., whether GUP really has the proposed form of $N$-dependence, and whether black hole complementarity is indeed correct. \\  \newline

{\tt Preprint Number:  YITP-14-61.}

\newpage
\addtocounter{section}{1}
\section* {\large{\textsf{1. Information Loss and Firewall: The Role of Quantum Mechanics}}}

The nature of Hawking radiation \cite{Hawking1, Hawking2} remains a puzzle 40 years after its conception --- does the radiation carry any information about matter that falls into the black hole, perhaps via subtle quantum entanglement? If not, gravitational collapse of a pure state seems to lead to a mixed state after the black hole evaporates away [see however, \cite{myers, arzano}], which has been argued to be a violation of unitarity in quantum mechanics. This is the so-called \emph{information loss paradox}, although not everyone agrees that this is a problem \cite{Wald1}. 
If one assumes that purity is recovered at the end by maximally entangling the late time Hawking radiation to the early ones \cite{page1, page2}, we have to demand consistency between general relativity and unitary quantum theory \cite{kn:stu}. However, this leads to some inconsistencies \cite{Yeom:2008qw1,Yeom:2008qw2,Yeom:2008qw0}.
In an attempt to resolve these conflicts, AMPS introduced a ``firewall'' at the black hole boundary, which prevents any act of probing the interior of the black hole horizon \cite{amps, apologia}, by incinerating any infalling observer [see also \cite{sam}]. If there exists a firewall, then it descends onto the horizon 
when the black hole lost about half of its Bekenstein-Hawking entropy \cite{page1, page2}, at which point it can still be quite large, and therefore has negligible curvature at the horizon. Thus it would seem that if firewall exists, our quest for quantum gravity has led us to a theory that does \emph{not} reduce back to quantum field theory on curved spacetime at the energy scale that it should have been valid. 

The black hole information loss paradox and the firewall controversy is a manifestation of the incompatibility between quantum physics and general relativity. While firewall proponents are quick to embrace unitarity, some have argued that perhaps we should closely examine our understanding of quantum mechanics [and quantum field theory]. Furthermore one should perhaps consider ``all possible histories'' that contribute to the Feynman path integral when discussing unitarity, and that unitarity \emph{is} preserved if we consider the fact that Alice doesn't always fall into the black hole in all ``branches'' of the wavefunction\footnote{The word ``branches'' suggests the Many-World Interpretation of quantum mechanics, but this is perhaps not necessary as one can phrase this in terms of decoherence. See however, \cite{Hsu0}.} \cite{Hsu1, Hsu2, Hsu3, Hollowood, SY, OS}. It also remains a possibility that quantum mechanics should be modified when gravitational effects are strong, or in the vicinity of trapped surfaces such as black holes. One possibility is to allow for non-locality \cite{Giddings}. Yet another modification to quantum mechanics is the so-called ``Generalized Uncertainty Principle'' [GUP], which generalizes the usual Heisenberg's Uncertainty Principle \cite{1}. However, the implication of GUP for information loss problem has not yet been well-studied\footnote{See however, the work by Itzhaki \cite{Itzhaki}, in which it was claimed that a large black hole cannot be described by means of local field theory even at macroscopic distances, because ``near the horizon the limitations on spacetime measurement are of the order of the black hole radius''. Indeed, although Itzhaki did not mention GUP, his Eq.(3) takes the form that we would recognize as that of GUP.}. 

In this work, we will start with a short review of GUP, and then investigate its implication for information loss paradox in the context of black hole complementarity principle \cite{kn:stu}, which proposed that quantum mechanics should only be consistent with causality. Namely, if Alice brings in a [localized] quantum state into the black hole and the exterior observer Bob recovers the information in the Hawking radiation, the apparent cloning of quantum information is actually allowed since these two observers are out of causal contact and cannot compare notes. One way to interpret this is to say that there is \emph{no} actual cloning --- the interior degrees of freedom are the same as the exterior ones, and therefore Bob who stays outside the black hole, can describe physics unitarily without having to care about what happens inside the black hole. Indeed, quantum mechanics only concerns what is actually [in principle] observable\footnote{A similar idea that information is physical only if it can be decoded has recently been proposed by Harlow and Hayden \cite{kn:HH, suss-2} in an attempt to resolve the firewall paradox. See also some follow-up works in \cite{OU,OMC,OC}.}. In order for the complementarity principle to be self-consistent, it is necessarily that Bob cannot communicate with Alice \emph{under any circumstances}. For example, after collecting Hawking radiation for a long time [so that Alice's message already comes out, due to unitarity requirement], Bob could jump into the black hole and attempts to receive a message sent by Alice. If this is possible, then Bob could have in his possession two copies of the same [arbitrary] quantum states, in violation of the No-Cloning Theorem of quantum information. Checking this consistency was an important test of the complementarity principle \cite{ST, kn:hp}, and it is interesting to see what happens if GUP is taken into account. [Previous study of black hole complementarity in the context of GUP, albeit in a different context, can be found in, e.g., \cite{MM}.]

\addtocounter{section}{2}
\section* {\large{\textsf{2. The Generalized Uncertainty Principle and Black Hole Physics}}}

One of the most important features of quantum mechanics is the fact that there is a fundamental limit on the precision with which some pairs of observables can be measured. This is the Heisenberg's Uncertainty Principle familiar to physics undergraduates, obeyed by position $x$ and momentum $p$: 
\begin{equation}\label{HUP}
\Delta x \Delta p \geqslant \frac{\hbar}{2}.
\end{equation}
This standard uncertainty principle is of course deduced under the assumption that the background spacetime is Minkowskian. In the presence of strong gravity, one expects modification to the uncertainty principle. Such modification can be obtained by quite general considerations of quantum mechanics and gravity \cite{1, 2, 3, 4}, but it also has support from string theoretical considerations \cite{5, 6, 7, 8, 9}. The result is the Generalized Uncertainty Principle [GUP], given by
\begin{equation}\label{GUP}
\Delta x\Delta p \geqslant \frac{1}{2}\left[\hbar + \alpha L_{\text{P}}^2\frac{(\Delta p)^2}{\hbar}\right],
\end{equation}
where $L_{\text{P}} = \sqrt{G\hbar/c^3}$ is the Planck length and $\alpha$ is a dimensionless parameter\footnote{Phenomenologically we assume $\alpha > 0$; in string theoretical derivations of GUP, $\alpha$ is essentially the Regge slope parameter $\alpha' > 0$, related to the string length $\lambda_s$ by $\lambda_s^2 = \hbar \alpha'$.} of order unity [however, see more discussions below]. In order to be consistent with Eq.(\ref{HUP}), our expression of GUP differs from \cite{1} by a factor of 1/2. 

It turns out that the usual Heisenberg's Uncertainty Principle allows \emph{heuristic} ``derivation'' of the Hawking temperature of Schwarzschild black hole in asymptotically flat spacetime \cite{ACS}. In this work we will only deal with 4-dimensional spacetimes for simplicity.

The Schwarzschild metric is
\begin{equation}
g[\text{Sch}] = -\left(1-\frac{2GM}{c^2r}\right)dt^2 + \left(1-\frac{2GM}{c^2r}\right)^{-1}dr^2 + r^2d\Omega^2, 
\end{equation}
where $d\Omega^2$ is the standard metric on a 2-sphere. The Schwarzschild event horizon is at $r_{\text{h}} = 2GM/c^2$.

Heisenberg's Uncertainty Principle then yields -- if we identify $\Delta x \sim r_{\text{h}}$ -- the following approximation:
\begin{equation}\label{1}
\Delta p \sim \frac{\hbar}{2\Delta x} \sim \frac{\hbar}{2r_{\text{h}}} = \frac{\hbar c^2}{4GM}.
\end{equation}
The reason for identifying $\Delta x$ with $r_{\text{h}}$ is as follows: we imagine a wave packet of quantum particle in the black hole spreading over spatial distance of $\Delta x$, if this distance is of the same order as the size of the black hole characterized by the horizon, then there is a chance for the particle to be found \emph{outside} the horizon as emitted particle. Of course, this argument being heuristic, overlooks the fact that the $r$ coordinate in the Schwarzschild spacetime is only an area radius and thus does not correspond to physical distance, and also the fact that the interior of the black hole is not static and $r$ plays the role of time there\footnote{We remark that this heuristic derivation only seems to work with Schwarzschild black hole, once there are more length scales involved, such as in the case of charged black holes, it is no longer obvious how to make such a heuristic argument work.}. 

From Eq.(\ref{1}), the uncertainty in the energy of photons emitted during Hawking evaporation is then identified with
\begin{equation}
\Delta pc \sim \frac{\hbar c^3}{4GM} = 2\pi \left(\frac{\hbar c^3}{8\pi GM}\right) = 2\pi T_{\text{BH}},
\end{equation}
where the Boltzmann constant $k_B$ has been set to unity, and
\begin{equation}
T_{\text{BH}} = \frac{\hbar c^3}{8\pi GM},
\end{equation}
which is the Hawking temperature of the Schwarzschild black hole. That is to say, the heuristic ``derivation'' gives the Hawking temperature up to a ``calibration factor'' of $2\pi$.
Now we repeat the heuristic argument of Hawking evaporation by replacing the Heisenberg's Uncertainty Principle with the GUP, and obtain [upon inserting the ``calibration factor''] the modified Hawking temperature \cite{ACS}
\begin{equation}\label{ACS}
T_{\text{GUP}} = \frac{M}{4\pi}\left(1-\sqrt{1-\frac{M_p^2}{\alpha M^2}}\right).
\end{equation}
From Eq.(\ref{ACS}) onward, we have set $c=\hbar=k_B=1$. Therefore $G=M_{\text{P}}^{-2}$, where $M_{\text{P}}$ is the Planck mass. Also, in this unit, the Planck length is $L_{\text{P}} = M_{\text{P}}^{-1}$. 

The generalized uncertainty principle in Eq.(\ref{GUP}) can then be written as
\begin{eqnarray}
\Delta x \Delta p \geqslant  \frac{1}{2} \left( 1 + \alpha \frac{\Delta p^{2}}{M_{\mathrm{P}}^{2}} \right).
\end{eqnarray}
If $\alpha \rightarrow 0$, then this recovers the Heisenberg's uncertainty principle.

From the GUP, we can derive useful relations. For example, the bound on the momentum uncertainty:
\begin{eqnarray}
\frac{M_{\mathrm{P}}^{2}}{\alpha} \Delta x \left( 1 - \sqrt{1 - \frac{\alpha}{M_{\mathrm{P}}^{2} \Delta x^{2}}} \right) \leqslant \Delta p \leqslant \frac{M_{\mathrm{P}}^{2}}{\alpha} \Delta x \left( 1 + \sqrt{1 - \frac{\alpha}{M_{\mathrm{P}}^{2} \Delta x^{2}}} \right).
\end{eqnarray}
From this, we can observe two important points, namely
\begin{itemize}
\item[(1)] The square root imposes a bound on the position uncertainty: $\Delta x \geqslant \Delta x_{\mathrm{min}}$, where
\begin{eqnarray}
\Delta x_{\mathrm{min}} := \frac{\sqrt{\alpha}}{M_{\mathrm{P}}}=\sqrt{\alpha}L_{\text{P}}.
\end{eqnarray}
Therefore, GUP naturally incorporates a \textit{minimal length} in the theory. [For a review on minimal length in quantum gravity, see \cite{sabine}.] 
\item[(2)] The maximum energy associated to probing a distance $\Delta x$ is
\begin{eqnarray}
\Delta E_{\mathrm{max}} := \frac{M_{\mathrm{P}}^{2}}{\alpha} \Delta x|_{\text{min}} \left( 1 - \sqrt{1 - \frac{\alpha}{M_{\mathrm{P}}^{2} \Delta x|_{\text{min}}^{2}}} \right).
\end{eqnarray}
In other words, if there is a position uncertainty $\Delta x$, then there exists an associated energy uncertainty, at most $\Delta E_{\mathrm{max}}$. If $\Delta x$ is of the order of the black hole radius, then $\Delta E_{\mathrm{max}}$ corresponds to the [modified] Hawking temperature \cite{ACS}.
\end{itemize}

We remark that it is possible for $\alpha$ to depend on the number of species that contribute to the Hawking radiation such that
$\alpha \propto N$, where $N$ is the number of species [e.g., the number of massless scalar fields that contribute to the Hawking radiation] \cite{Dvali:2007hz1,Dvali:2007hz2,Dvali:2007hz3}. 
The value of this parameter can in principle be somewhat constrained by observations \cite{DV}.
In order for the black hole to remain semi-classical, we require that the black hole should be much larger than the minimum length: $r_{\mathrm{h}} \gg \sqrt{N}/M_{\mathrm{P}}$. While such an $N$-dependence is not rigorously proven, as we will later show, GUP provides a good framework to be consistent with black hole complementarity precisely if such $N$-dependence is allowed, whereas the usual argument that depends only on the Heisenberg's uncertainty relation fails.

Finally, we comment on some interesting limits:
\begin{itemize}
\item[(1)] $\alpha \rightarrow 0$ limit, as mentioned before, recovers the Heisenberg's uncertainty principle:
\begin{eqnarray}
\frac{1}{2 \Delta x} \leqslant \Delta p \leqslant \frac{2M_{\text{P}}^2}{\alpha} \Delta x \to \infty.
\end{eqnarray}
\item[(2)] $\alpha \to M_{\mathrm{P}}^{2} \Delta x^{2}$ limit:
\begin{eqnarray}
\Delta p \to \frac{M_{\mathrm{P}}^{2}}{\alpha} \Delta x.
\end{eqnarray}
This is the optimal limit, that is, given any fixed $\Delta p$, the smallest $\Delta x$ is obtained in this limit.
\end{itemize}

\addtocounter{section}{3}
\section* {\large{\textsf{3. Black Hole Complementarity and Generalized Uncertainty Principle}}}

In order for the complementarity principle to be a correct description, one has to check whether it is possible for the infalling Alice to send her quantum bit to Bob who falls into the black hole at a later time, after he has obtained a copy of the same bit from the Hawking radiation. As we will review later, the longer Bob waits outside, the shorter the available time Alice has to send her bit before she crashes into the singularity [or whatever replaces the singularity in a full quantum gravity theory]. We first review this quantum bit duplication thought experiment in the standard picture before applying GUP\footnote{In the following, we assume the standard local QFT and quantum entanglement to hold as per the usual requirement in black hole complementarity. This is however, not a trivial statement. In fact, quantum entanglement may exhibit novel features on curved spacetimes \cite{FSM}.}.

We first prepare an entangled spin pair $\left|a\right\rangle$ and $\left|b\right\rangle$. If $\left|a\right\rangle$ is in the up state, then $\left|b\right\rangle$ is in the down state, and vice versa. We assume that there is an in-going observer Alice, $\mathcal{A}$, who brings $\left|a\right\rangle$ into the black hole. After a certain time, Alice sends a signal regarding the spin $\left|a\right\rangle$ in the ``out-going'' [being inside the black hole, this signal cannot propagate out to the null infinity] direction.

Meanwhile, $\left|b\right\rangle$ is outside the horizon. We assume that there is another observer Bob, $\mathcal{B}$, who is outside the event horizon and measures $\left|b\right\rangle$. Therefore, Bob knows the state of $\left|b\right\rangle$, whether it is up or down. After the Page time or the information retention time \cite{page1,page2}, $t_{\text{info}} \sim GM^3/M_\text{P}^2$, Hawking radiation emits the information of $\left|a\right\rangle$: we call this $\left|h\right\rangle$. Then Bob can measure $\left|h\right\rangle$ outside the horizon. By comparing with $\left|b\right\rangle$, Bob notices that this information is in fact $\left|a\right\rangle$. [In a ``realistic experiment'', this should be repeated many times. The correlation between $\left|h\right\rangle$ and $\left|b\right\rangle$ will then become more obvious.]

Finally, Bob falls into the black hole. If Alice sends a signal of $\left|a\right\rangle$ fast enough, Bob can eventually see $\left|a\right\rangle$ on his trip toward the [future spacelike] singularity. Then, he knows that $\left|a\right\rangle$ is indeed the original information by comparing with $\left|b\right\rangle$. [Again, in a ``realistic experiment'', this should be repeated to ensure the clear correlation between $\left|a\right\rangle$ and $\left|b\right\rangle$].

If all of these processes are possible, then Bob sees a duplication of information $\left|a\right\rangle$, which contradicts the No-Cloning Theorem. Therefore, this will be inconsistent with the assumptions of black hole complementarity.

To make this thought experiment possible, we need two requirements:
\begin{itemize}
\item[(1.)] The observer $\mathcal{B}$ should fall into the black hole \textit{after} the Page time [information retention time] $t_{\text{info}} \sim GM^{3}/M_{\text{P}}^2$.
\item[(2.)] The observer $\mathcal{A}$ should successfully send a signal to the observer $\mathcal{B}$ before either of them crashes into the singularity.
\end{itemize}

After a simple calculation \cite{ST}, we can show that the observer $\mathcal{A}$ should send a signal within the time interval of
\begin{eqnarray}\label{time}
\Delta \tau \simeq r_{\mathrm{h}} \exp \left( - \frac{\Delta t}{r_{\mathrm{h}}} \right), 
\end{eqnarray}
where $r_{\mathrm{h}}\sim GM$ is the black hole horizon, $\Delta \tau$ is Alice's proper time available to send message, and $\Delta t$ is the Schwarzschild coordinate time delay between Alice and Bob. Here it is evident that the longer Bob stays outside collecting Hawking radiation, the less time Alice has to send her message. 

To send a bit of quantum information within the time $\Delta \tau$ requires some energy $\Delta E$, and we have to rely on the uncertainty relation. Indeed, to send a bit of information between $\Delta \tau$, one needs, with the Heisenberg's uncertainty relation $\Delta E \Delta \tau \sim 1$, 
\begin{eqnarray}
\Delta  E \simeq r_{\mathrm{h}}^{-1} \exp \left( + \frac{\Delta t}{r_{\mathrm{h}}} \right).
\end{eqnarray}
That is to say, to send message in a short time interval $\Delta \tau$ requires the message to be encoded in high enough energy [since energy is exponential in $\Delta t$].
The longer Bob waits, the shorter the time Alice has, and the larger the energy she needs to send the message. Eventually the required energy becomes greater than that of the black hole itself, i.e., $\Delta E > M$, and such a message sending act would become impossible. 

Requiring that $\Delta \tau < M^{-1}$ also implies that 
\begin{equation}
\Delta t \lesssim r_{\mathrm{h}} \log{\left(\frac{M r_{\mathrm{h}}}{M_{\text{P}}^2}\right)} \sim GM \log \left(\frac{M}{M_{\text{P}}}\right).
\end{equation}
The time scale $GM \log M/M_{\text{P}}$ is known as the \emph{scrambling time} \cite{suss1, suss2}. 
The consistency condition for complementarity principle to hold -- that is, Alice \emph{fails} to send message -- is thus 
\begin{equation}
\Delta t \gtrsim  GM \log \left(\frac{M}{M_{\text{P}}}\right). 
\end{equation}

Since the information retention time $t_\text{info}$ is the order of black hole lifetime, which is $GM^3/M_{\text{P}}^2$, we see that for a young black hole [before the turnover of the Page curve \cite{page1, page2}], complementarity is completely safe, since 
\begin{equation}
t_\text{info} \sim \frac{GM^3}{M_{\text{P}}^2} \gg GM \log \frac{M}{M_{\text{P}}} = GM \log (GMM_\text{P}). 
\end{equation}
However, one could use an old black hole to perform the thought experiment. Then, Bob already has in his possession more than half of the Hawking radiation before Alice jumps into the black hole with her bit. This bit will come out fairly quickly and complementarity seems just barely safe \cite{kn:hp}. 

However, sending a bit of information becomes \emph{possible even for young black holes} if we introduce a large number of scalar fields. The idea is this: as we have seen, the important time scale involved here is the black hole lifetime, which goes proportional to $M^3$. The lifetime itself is of course, controlled by the number of species of particle emitted in the Hawking radiation. So what happens if we have large number of particle species?

More specifically, the differential equation governing the evaporation rate of a neutral black hole is
\begin{equation}\label{massloss0}
\frac{dM}{dt} = - C \sigma T_{\text{BH}}^4, 
\end{equation}
which is just the familiar Stefan-Boltzmann law, with $a=\pi^2/(15\hbar^3)$ being the radiation constant. 
For a large black hole, only massless particle emission is important since the black hole is too cold to emit substantial amount of massive particles.
The quantity $\sigma$ thus denotes the area of the emitting surface, which is the surface that corresponds to the photon orbit. This is because only particles that have enough energy can escape the effective potential barrier, with local maximum at the photon orbit [see Fig.(6.5) of \cite{Wald}].
The constant $C$ depends on the number of species of massless particles. 
This grey-body factor usually only contributes to $O(1)$ correction since the number of [known] massless particles is $O(1)$ \cite{page0} [for charged black hole, one could see how the black hole lifetime depends on the number of particle species in, e.g., Fig.(4) of \cite{kn:HW}]. However, the lifetime can be considerably shortened if one considers sufficiently large $N$ number of massless particles --- it is of the order $M^3/N$.

One way to explore the consequence of large number of scalar field is the so-called  ``large $N$ rescaling''  \cite{Yeom:2008qw0} of evaporating black holes, which we will now explain.

Let us consider a system in which a semi-classical black hole is formed via collapse of a single scalar field, but evaporates by emitting $N$ scalar fields. Then the black hole satisfies the semi-classical Einstein equations:
\begin{eqnarray}
G_{\mu\nu} = \frac{8\pi G}{c^4} \left(T_{\mu\nu} + \hbar N \langle \mathcal{T}_{\mu\nu}\rangle \right)
\end{eqnarray}
up to order $\hbar$ , where $\langle \mathcal{T}_{\mu\nu}\rangle$ denotes the energy-momentum contribution from Hawking radiation. 
By choosing the unit $c=N=\hbar=1$, we can in principle obtain a solution of the equations. Having obtained the solution, the idea of large $N$ rescaling is to \emph{not} fix $\hbar$, but only $N\hbar=1$. If $N$ increases, then conversely $\hbar$ decreases, and hence in general, the units of length, time, and mass decrease accordingly by a factor of $\sqrt{N}$, and so we obtained a family of solutions corresponding to different number of fields and values of $\hbar$. In other words, the physical size increases in Planck units \cite{Yeom:2008qw1,Yeom:2008qw2,Yeom:2008qw0,Yeom:2008qw3}. While it may seem strange to vary $\hbar$, the idea is that to check the validity of complementarity principle, it is enough to invalidates it with one counterexample, even if the universe in which complementarity fails has different value of $\hbar$ than our own. After all, it does not seem plausible that the fundamental solution to the information loss problem should depend on the exact value of $\hbar$ as long as it is finite and nonzero.  [It is also possible that in our universe, $\hbar$ can be varied as a field, i.e. it may be spacetime dependent \cite{0454}. ]

Under large $N$ rescaling scheme, the time difference and the mass should be rescaled as follows:
\begin{eqnarray}
\Delta x \rightarrow \Delta x':=\sqrt{N} \Delta x, \;\;\;\;  M \rightarrow M':=\sqrt{N} M.
\end{eqnarray}
In other words, the family of black hole solutions with mass $\sqrt{N}M$ emitting $N$ species of massless scalar particles, has the same causal structure, in the sense that the ratio of lifetime over horizon size remains the same. That is, if we change the number of fields, the said ratio is maintained if we multiply the mass by a factor of $\sqrt{N}$ [in 4-dimensions]. Note that Hawking radiation $T_{\text{BH}} \propto M^{-1}$, so that under rescaling, $T_{\text{BH}}$ becomes smaller with larger $N$, but Hawking radiation is still effective due to compensation from the large number of species of scalar particles.

Taking large $N$ rescaling into consideration, the required energy $\Delta E'$ for successful duplication experiments becomes 
\begin{eqnarray}
\Delta E' \simeq \frac{1}{\sqrt{N} \Delta \tau}.
\end{eqnarray}
The duplication experiment cannot be carried out if 
\begin{eqnarray}
\frac{1}{\sqrt{N} \Delta \tau} \gtrsim \sqrt{N} M.
\end{eqnarray}
In other words, if we do have sufficiently large number of scalar fields, such that
\begin{eqnarray}
N \gtrsim \frac{1}{M \Delta \tau},
\end{eqnarray}
then the observation of the duplication of information is allowed. Note that here $M$ and $\Delta \tau$ are the mass and the time difference that are measured in the $N=1$ case.

This argument can be further strengthened if we consider two following points. Firstly, we can apply the similar argument for not only the information retention time, but also the scrambling time. Then the required number of scalar fields can be reduced \cite{Yeom:2008qw0}. Secondly, at least in two-dimensional spacetimes, even with the information retention time, the required number of scalar fields can be reduced to a ``reasonable number'' that can ``surely'' be allowed by string theory \cite{Kim:2013fv}.

So it seems that black hole complementarity will be in trouble if there exists enough massless scalar fields. 
However, the calculations above depend crucially on the validity of the Heisenberg's uncertainty relation, all the way to ``near''-singularity region [``near'' in the temporal direction]. If quantum mechanics does receive correction due to gravitational effects, one should re-check the calculations by using GUP. 

A novel feature of GUP is the existence of a minimum length, that is,
$\Delta x \geqslant \Delta x_{\mathrm{min}} = \sqrt{\alpha}/M_{\text{P}}$. This translates into uncertainty in time upon dividing it by $c=1$, and we see that
sending information to Bob would require a minimum time
\begin{eqnarray}\label{GUPtime}
t_{\text{min}} \sim GM \log \frac{GM M_{\mathrm{P}}}{\sqrt{\alpha}}.
\end{eqnarray}
To be more explicit, Eq.(\ref{GUPtime}) is derived by equating the minimal length $\Delta x_{\mathrm{min}}$ [upon dividing by $c=1$] with Eq.(\ref{time}).

Assuming the GUP, if $\alpha$ does depend on $N$ such that $\alpha \propto N$, as proposed in \cite{Dvali:2007hz1,Dvali:2007hz2,Dvali:2007hz3}, then we see that $\Delta x/\Delta x_{\mathrm{min}}$ is invariant up to the choice of $N$, and hence even after the large $N$ rescaling,
\begin{eqnarray}
\frac{\Delta x}{\Delta x_{\mathrm{min}}} \geqslant 1
\end{eqnarray}
holds. Therefore, if $\alpha$ depends on $N$ \emph{in an appropriate way}, then the generalized uncertainty principle would again prevent the quantum cloning of information. 

To be more specific, the original consistency relation required for the black hole complementarity 
\begin{equation}\label{consistency}
\frac{GM^3}{M_\text{P}^2} \gg GM \log\left( \frac{GM M_{\text{P}}}{\sqrt{\alpha}}\right)
\end{equation}
would become, under large $N$ rescaling,
\begin{equation}
\frac{G(\sqrt{N}M)^3}{M_{\text{p}}^2 N} \gg G(\sqrt{N}M)  \log \left(\frac{G\sqrt{N} M M_{\text{P}}}{\sqrt{\alpha}}\right),
\end{equation}
where we have divided by $N$ on the left hand side of the inequality because the lifetime should decrease by a factor of $1/N$, as previously explained.
Thus, if $\alpha \sim N$, then  
\begin{equation}
\frac{G(\sqrt{N}M)^3}{M_{\text{p}}^2 N} \gg  G (\sqrt{N}M)  \log (G M M_{\text{P}}),
\end{equation}
which has the same structure as the original consistency relation inequality Eq.(\ref{consistency}), as dependence of $N$ now drops out entirely. Therefore the inequality trivially holds for all values of $N$.

Even though such an $N$-dependence of $\alpha$ is far from obvious, this at least demonstrates that a correction to quantum mechanics in the form of generalized uncertainty principle can help to make complementarity principle works even under large $N$ rescaling. On the other hand, an opponent of black hole complementarity could claim the other way round that even with GUP, complementarity can still get into trouble if $\alpha$ is not $N$-dependent, or its $N$-dependence does not take the ``correct'' form. It thus remains an interesting and important question as to whether:
\begin{itemize}
\item[(1)] GUP should have the proposed form of $N$-dependence, and
\item[(2)] Black hole complementarity principle is correct under large $N$ rescaling.
\end{itemize}
Our objective in this work is only to point out the implication of GUP to black hole ccomplementarity, and therefore we leave these issues for future research.

\addtocounter{section}{4}
\section* {\large{\textsf{4. Discussion}}}

In this work, we re-examined black hole complementarity by considering a thought experiment in which Alice tries to send information to Bob, who first collected Hawking radiation in the exterior region before jumping into the black hole himself. Although such information sending cannot be performed if one considers the usual Heisenberg's Uncertainty Principle, it seems that by introducing sufficient numbers of massless scalar fields that contribute to the Hawking radiation, such an act now becomes possible. Whether GUP can save the complementarity principle depends on whether the GUP correction term is $N$-dependent in an appropriate way, where $N$ is the number of massless scalar fields contributing to the Hawking flux. Therefore, if black hole complementarity is indeed a correct principle for black hole physics, then this provides a guide for us to understand how GUP should behave, and in turn this may shed some insights on quantum gravity.

Of course one must also consider the unfortunate possibility that complementarity principle may not be correct [see also \cite{Yeom:2008qw1}, and of course \cite{amps, apologia}]. Regardless of the status of black hole complementarity, at least one application of GUP \cite{Itzhaki} to the exterior observer seems to suggest that spacetime measurement around a black hole has uncertainty of the order of the horizon radius. If this is correct, then it is tempting, though perhaps somewhat conjectural, to give some further thoughts on this observation as follows: we can consider an exterior observer Bob, together with the black hole, to be in a coherent quantum system, provided he does not ``disturb the black hole''. The wavefunction evolution is completely unitary. However, Alice who falls into the black hole and gets to probe the black hole interior, corresponds to a \emph{particular} history of the wavefunction [in terms of Many-World language, a particular Everett's branch --- there exist other branches, other histories, in which Alice missed the black hole completely], as advocated in \cite{Hsu2}. GUP may thus provide a natural context to reconcile unitarity as perceived by an exterior observer who remains coherent with [macroscopic] superpositions of black hole states, and the infalling observer who does not experience anything special at the horizon, fully in agreement with quantum field theory on curved spacetime in its regime of validity. The details of this argument will be pursued elsewhere. 

One remaining issue to discuss is that GUP has been invoked to argue for the existence of black hole remnants, i.e., black holes don't completely evaporate since Hawking evaporation eventually stops as the hole becomes Planckian in size \cite{ACS, IMN}. In view of the usual objections against remnants [e.g. infinite pair-production], one naturally wonders if this means that GUP itself is somewhat problematic. Then again, remnants may not be as problematic as usually thought \cite{sabinesmolin}. We leave this question open for future considerations.

\addtocounter{section}{1}
\section*{\large{\textsf{Acknowledgement}}}
Pisin Chen and Yen Chin Ong thank Taiwan's National Center for Theoretical Sciences [NCTS], Taiwan National Science Council [NSC], and the Leung Center for Cosmology and Particle Astrophysics [LeCosPA] of National Taiwan University for support. Dong-han Yeom is supported by the JSPS Grant-in-Aid for Scientific Research (A) No. 21244033. Yen Chin Ong also thanks Keisuke Izumi for useful discussions.



\begin{thebibliography}{99}

\bibitem{Hawking1}
Stephen Hawking, ``Black Hole Explosions?'', Nature \textbf{248} (1974) 30.

\bibitem{Hawking2}
Stephen Hawking, ``Particle Creation by Black Holes'', Commun. Math. Phys. \textbf{43} (1975) 199.

\bibitem{myers}
Robert C. Myers, ``Pure States Don't Wear Black'', Gen. Rel. Grav. \textbf{29} (1997) 1217, \href{http://arxiv.org/abs/gr-qc/9705065v1}{[arXiv:gr-qc/9705065]}.

\bibitem{arzano}
Michele Arzano, ``Purity is Not Eternal at the Planck Scale'', Phys. Rev. D 90 (2014) 024016, \href{http://arxiv.org/abs/1403.6457}{[arXiv:1403.6457 [hep-th]]}.


\bibitem{Wald1}
Robert M. Wald, ``The Thermodynamics of Black Holes'', 
Living Rev. Relativity \textbf{4} (2001) 6, \href{http://www.livingreviews.org/lrr-2001-6}{accessed: August 15, 2014}.

\bibitem{page1}
Don N. Page, ``Average Entropy of a Subsystem'', Phys. Rev. Lett. \textbf{71} (1993) 1291, \href{http://arxiv.org/abs/gr-qc/9305007}{[arXiv:gr-qc/9305007]}.

\bibitem{page2}
Don N. Page, ``Time Dependence of Hawking Radiation Entropy'', JCAP \textbf{1309} (2013) 028 , \href{http://arxiv.org/abs/1301.4995}{[arXiv:1301.4995 [hep-th]]}.

\bibitem{kn:stu}
Leonard Susskind, Larus Thorlacius, John Uglum, ``The Stretched Horizon and Black Hole Complementarity'', Phys. Rev. D \textbf{48} (1993) 3743, \href{http://arxiv.org/abs/hep-th/9306069}{[arXiv:hep-th/9306069]}.

 \bibitem{Yeom:2008qw1}
  Dong-han Yeom, Heeseung Zoe,
  ``Constructing a Counterexample to the Black Hole Complementarity'',
  Phys.\ Rev.\  D {\bf 78} (2008) 104008, \href{http://arxiv.org/abs/0802.1625}{
 [arXiv:0802.1625 [gr-qc]]}.

    
\bibitem{Yeom:2008qw2} 
  Sungwook E. Hong, Dong-il Hwang, Ewan D. Stewart, Dong-han Yeom,
  ``The Causal Structure of Dynamical Charged Black Holes,''
  Class.\ Quant.\ Grav.\  {\bf 27} (2010) 045014, \href{http://arxiv.org/abs/0808.1709}{
  [arXiv:0808.1709 [gr-qc]]}.
 
\bibitem{Yeom:2008qw0}  
	Dong-han~Yeom and Heeseung~Zoe,
  ``Semi-Classical Black Holes with Large N Re-Scaling and Information Loss Problem,''
  Int.\ J.\ Mod.\ Phys.\ A {\bf 26} (2011) 3287, \href{http://arxiv.org/abs/0907.0677}{
  [arXiv:0907.0677 [hep-th]]}.




\bibitem{amps}
Ahmed Almheiri, Donald Marolf, Joseph Polchinski, James Sully, ``Black Holes: Complementarity or Firewalls?'', JHEP \textbf{1302} (2013) 062, \href{http://arxiv.org/abs/1207.3123}{[arXiv:1207.3123 [hep-th]]}.

\bibitem{apologia}
Ahmed Almheiri, Donald Marolf, Joseph Polchinski, Douglas Stanford, James Sully,
``An Apologia for Firewalls'', JHEP \textbf{1309} (2013) 018, \href{http://arxiv.org/abs/1304.6483}{[arXiv:1304.6483 [hep-th]]}.

\bibitem{sam}
Samuel L. Braunstein, Stefano Pirandola, and Karol \.Zyczkowski, ``Better Late than Never: Information Retrieval from Black Holes'', Phys. Rev. Lett. \textbf{110} (2013) 101301, \href{http://arxiv.org/abs/0907.1190}{[arXiv:0907.1190 [quant-ph]]}.




\bibitem{Hsu0}
Stephen D. H. Hsu, David Reeb, ``Black Holes, Information and Decoherence'', Phys. Rev. D \textbf{79} (2009) 124037, \href{http://arxiv.org/abs/0903.2258}{[arXiv:0903.2258 [gr-qc]]}.

\bibitem{Hsu1}
Stephen D. H. Hsu, ``The Black Hole Information Paradox and Macroscopic Superpositions'', J. Phys. Conf. Ser. \textbf{222} (2010) 012037, \href{http://arxiv.org/abs/1003.5382}{[arXiv:1003.5382 [gr-qc]]}.

\bibitem{Hsu2}
Stephen D. H. Hsu, ``Macroscopic Superpositions and Black Hole Unitarity'', \href{http://arxiv.org/abs/1302.0451}{[arXiv:1302.0451 [hep-th]]}.

\bibitem{Hsu3}
Stephen D. H. Hsu, ``Factorization of Unitarity and Black Hole Firewalls'', \href{http://arxiv.org/abs/1308.5686}{[arXiv:1308.5686 [hep-th]]}.

\bibitem{Hollowood}
Timothy J. Hollowood, ``Schrodinger's Cat and the Firewall'', \href{http://arxiv.org/abs/1403.5947}{[arXiv:1403.5947 [hep-th]]}.

\bibitem{SY}
Misao Sasaki, Dong-han Yeom, ``Thin-Shell Bubbles and Information Loss Problem in Anti-de Sitter Background'', \href{http://arxiv.org/abs/1404.1565}{[arXiv:1404.1565 [hep-th]]}.

\bibitem{OS}
Elias Okon, Daniel Sudarsky, ``The Black Hole Information Paradox and the Collapse of the Wave Function'', \href{http://arxiv.org/abs/1406.2011}{[arXiv:1406.2011 [gr-qc]]}.

\bibitem{Giddings}
Steven B. Giddings, ``Nonviolent Nonlocality'',  Phys. Rev. D \textbf{88} (2013) 064023, \href{http://arxiv.org/abs/1211.7070}{[arXiv:1211.7070 [hep-th]]}.

\bibitem{1} Ronald J. Adler, David I. Santiago, ``On Gravity and the Uncertainty Principle'', Mod. Phys. Lett. A \textbf{14} (1999) 1371, \href{http://arxiv.org/abs/gr-qc/9904026}{	[arXiv:gr-qc/9904026]}. 


\bibitem{Itzhaki}
Nissan Itzhaki, ``Black Hole Information vs. Locality'', Phys. Rev. D \textbf{54} (1996) 1557, \href{http://arxiv.org/abs/hep-th/9510212}{[arXiv:hep-th/9510212]}.



\bibitem{kn:HH}
Daniel Harlow, Patrick Hayden, ``Quantum Computation vs. Firewalls'', JHEP \textbf{06} (2013) 085, \href{http://arxiv.org/abs/1301.4504}{[arXiv:1301.4504
[hep-th]]}.

\bibitem{suss-2}
Leonard Susskind, ``Black Hole Complementarity and the Harlow-Hayden Conjecture'', \href{http://arxiv.org/abs/1301.4505}{[arXiv:1301.4505 [hep-th]]}.

\bibitem{OU}
Jonathan Oppenheim, William G. Unruh, ``Firewalls and Flat Mirrors: An Alternative to the AMPS Experiment which Evades the Harlow-Hayden Obstacle'', JHEP \textbf{1403} (2014) 120 , \href{http://arxiv.org/abs/1401.1523}{[arXiv:1401.1523 [hep-th]]}.

\bibitem{OMC}
Yen Chin Ong, Brett McInnes, Pisin Chen, ``Why Hawking Radiation Cannot Be Decoded'', \href{http://arxiv.org/abs/1403.4886}{[arXiv:1403.4886 [hep-th]]}.

\bibitem{OC}
Yen Chin Ong, Pisin Chen, ``Charge Loss (or the Lack Thereof) for AdS Black Holes'', JHEP \textbf{06} (2014) 061, \href{http://arxiv.org/abs/1404.5215}{[arXiv:1404.5215 [gr-qc]]}.

\bibitem{ST}
Leonard Susskind, Larus Thorlacius, ``Gedanken Experiments Involving Black Holes'', Phys. Rev. D \textbf{49} (1994) 966, \href{http://arxiv.org/abs/hep-th/9308100}{[arXiv:hep-th/9308100]}.

\bibitem{kn:hp}
Patrick Hayden, John Preskill, ``Black Holes as Mirrors: Quantum Information in Random Subsystems'', JHEP \textbf{0709} (2007) 120, \href{http://arxiv.org/abs/0708.4025}{[arXiv:0708.4025 [hep-th]]}.

\bibitem{MM}
Michele Maggiore, ``Black Hole Complementarity and the Physical Origin of the Stretched Horizon'', Phys. Rev. D \textbf{49} (1994) 2918, \href{http://arxiv.org/abs/hep-th/9310157}{[arXiv:hep-th/9310157]}.



\bibitem{2} Michele Maggiore, ``A Generalized Uncertainty Principle in Quantum Gravity'', Phys. Lett. B \textbf{304} (1993) 65, \href{http://arxiv.org/abs/hep-th/9301067}{[arXiv:hep-th/9301067]}.

\bibitem{3} Michele Maggiore, ``Quantum Groups, Gravity, and the Generalized Uncertainty Principle'', Phys. Rev. D \textbf{49} (1994) 5182, \href{http://arxiv.org/abs/hep-th/9305163}{[arXiv:hep-th/9305163]}. 

\bibitem{4} Fabio Scardigli, ``Generalized Uncertainty Principle in Quantum Gravity from Micro-Black Hole Gedanken Experiment'', Phys. Lett. B \textbf{452} (1999) 39, \href{http://arxiv.org/abs/hep-th/9904025}{[arXiv:hep-th/9904025]}. 

\bibitem{5} Gabriele Veneziano, ``A Stringy Nature Needs Just Two Constants'', Europhys. Lett. \textbf{2} (1986) 199. 

\bibitem{6} David J. Gross, Paul F. Mende, ``String Theory Beyond the Planck Scale'', Nucl. Phys. B \textbf{303} (1988) 407.

\bibitem{7} Daniele Amati, Marcello Ciafolini, Gabriele Veneziano, ``Can Spacetime be Probed Below the String Size?'', Phys. Lett. B \textbf{216} (1989) 41.

\bibitem{8} Kenichi Konishi, Giampiero Paffuti, Paolo Provero , ``Minimum Physical Length and the Generalized Uncertainty Principle in String Theory'', Phys. Lett. B \textbf{234} (1990) 276.

\bibitem{9} Edward Witten, ``Reflections on the Fate of Spacetime'', Phys. Today \textbf{49},  Apr. 24 (1996). 

\bibitem{ACS} Ronald J. Adler, Pisin Chen and David I. Santiago, ``The Generalized Uncertainty Principle and Black Hole Remnants'', Gen. Rel. Grav. \textbf{33} (2001) 2101, \href{http://arxiv.org/abs/gr-qc/0106080}{[arXiv:gr-qc/0106080]}.




\bibitem{sabine}
Sabine Hossenfelder, ``Minimal Length Scale Scenarios for Quantum Gravity'', 	Living Rev. Relativity \textbf{16} (2013) 2, \href{http://arxiv.org/abs/1203.6191}{[arXiv:1203.6191 [gr-qc]]}. 

\bibitem{Dvali:2007hz1}
  Gia Dvali, ``Black Holes and Large N Species Solution to the Hierarchy Problem'',
  Fortsch.\ Phys.\  {\bf 58} (2010) 528, \href{http://arxiv.org/abs/0706.2050}{
  [arXiv:0706.2050 [hep-th]]}
  
\bibitem{Dvali:2007hz2}
  Gia Dvali, Michele Redi, Sergey Sibiryakov, Arkady Vainshtein,
  ``Gravity Cutoff in Theories with Large Discrete Symmetries'',
  Phys.\ Rev.\ Lett.\  {\bf 101} (2008) 151603, \href{http://arxiv.org/abs/0804.0769}{
  [arXiv:0804.0769 [hep-th]]}. 
  
\bibitem{Dvali:2007hz3}
  Ram Brustein, Gia Dvali and Gabriele Veneziano, ``A Bound on the Effective Gravitational Coupling from Semiclassical Black Holes'',
  JHEP {\bf 0910} (2009) 085, \href{http://arxiv.org/abs/0907.5516}{
  [arXiv:0907.5516 [hep-th]]}.
  
\bibitem{DV}
Saurya Das, Elias C. Vagenas,
``Universality of Quantum Gravity Corrections'',
Phys. Rev. Lett. \textbf{101} (2008) 221301,
\href{http://arxiv.org/abs/0810.5333}{[arXiv:0810.5333 [hep-th]]}.
	
\bibitem{FSM}
Ivette Fuentes-Schuller, R. B. Mann, ``Alice Falls into a Black Hole: Entanglement in Non-Inertial Frames'', Phys. Rev. Lett. \textbf{95} (2005) 120404, \href{http://arxiv.org/abs/quant-ph/0410172}{[arXiv:quant-ph/0410172]}.
  
  
\bibitem{suss1}
Yasuhiro Sekino, Leonard Susskind, ``Fast Scramblers'', JHEP \textbf{0810} (2008) 065, \href{http://arxiv.org/abs/0808.2096}{[arXiv:0808.2096 [hep-th]]}. 

\bibitem{suss2}
Leonard Susskind, ``Addendum to Fast Scramblers'', \href{http://arxiv.org/abs/1101.6048v1}{[arXiv:1101.6048 [hep-th]]}.
  
\bibitem{Wald}
Robert Wald, \emph{General Relativity}, The University of Chicago Press, Chicago and London, 1984.



\bibitem{page0}
Don N. Page, ``Particle Emission Rates from a Black Hole: Massless Particles from an Uncharged, Nonrotating Hole'', Phys. Rev. D \textbf{13} (1976) 198.

\bibitem{kn:HW}
William A. Hiscock, Lance D. Weems,
``Evolution of Charged Evaporating Black Holes'', Phys. Rev. D \textbf{41} (1990) 1142.


\bibitem{Yeom:2008qw3} 
  Dong-han Yeom,
  ``Reviews and Perspectives on Black Hole Complementarity'',
  Int.\ J.\ Mod.\ Phys.\ CS {\bf 1} (2011) 311, \href{http://arxiv.org/abs/0901.1929}{[arXiv:0901.1929 [gr-qc]]}.

 
\bibitem{0454}
Sabine Hossenfelder, ``Gravity Can Be Neither Classical nor Quantized'', \href{http://arxiv.org/abs/1212.0454}{[arXiv:1212.0454 [gr-qc]]}. 

\bibitem{Kim:2013fv} 
  Wontae Kim, Bum-Hoon Lee, Dong-han Yeom,
  ``Black Hole Complementarity and Firewall in Two Dimensions'',
  JHEP {\bf 1305} (2013) 060, \href{http://arxiv.org/abs/1301.5138}{
  [arXiv:1301.5138 [gr-qc]]}.



\bibitem{IMN}
Maximiliano Isi, Jonas Mureika, Piero Nicolini, 
``Self-Completeness and the Generalized Uncertainty Principle'', 
JHEP {\bf 1311} (2013) 139, \href{http://arxiv.org/abs/1310.8153}{[arXiv:1310.8153 [hep-th]]}.



\bibitem{sabinesmolin}
Sabine Hossenfelder, Lee Smolin, ``Conservative Solutions to the Black Hole Information Problem'', Phys. Rev. D \textbf{81} (2010) 064009, \href{http://arxiv.org/abs/0901.3156}{[arXiv:0901.3156 [gr-qc]]}.











\end{thebibliography}
\end{document}